\documentclass{article}
\usepackage{amssymb,amsmath,amscd,epsf}
\begin{document}

\title{
\Large\sc
A remark on the
``Theory of neutrino oscillations''}

\author{
L.B. Okun\thanks{ITEP, Moscow, 117218, Russia e-mail: {\it
okun@heron.itep.ru} },  M.V. Rotaev\thanks{MIPT, Moscow, 141700,
Russia; ITEP, Moscow, 117218, Russia e-mail: {\it mrotaev@mail.ru}
},  M.G. Schepkin\thanks{ITEP, Moscow, 117218, Russia e-mail: {\it
schepkin@heron.itep.ru} },  I.S. Tsukerman\thanks{ITEP, Moscow,
117218, Russia e-mail: {\it zuckerma@heron.itep.ru} }~ }

\date{}

\maketitle

\begin{abstract}

It is shown that the derivation of standard oscillation phase in
the plane wave approximation without equal energy assumption is
fraught with inconsistencies.
\end{abstract}

 The literature on quantum mechanics of neutrino oscillations is vast and
controversial. Therefore, the words in ref \cite{Giunti} ``Let us
briefly describe the covariant derivation of the neutrino
oscillation probability in the plane wave approach'' attract
special attention. The derivation in ref. \cite{Giunti} is given
in six equations on one page and is based on the substitution $t
\simeq x = L$ in the phase $E_k t - p_k x$ for each of the three
mass states ($k=1,2,3$):
\begin{multline}
E_k t - p_k x \simeq (E_k - p_k)L = \frac{E_k^2-p_k^2}{E_k+p_k}L =
\frac{m_k^2}{E_k+p_k}L \simeq \frac{m_k^2}{2E}L \;\; ,
\end{multline}
where $L$ is the distance between the source and detector of
neutrinos. Our Eq.(1) reproduces the key equation of ref.
\cite{Giunti}, namely, Eq.(5).

It is further noticed in ref. \cite{Giunti}: ``It is important to
notice that Eq.(5) shows that the phase of massive neutrinos
relevant for the oscillations are independent from any assumptions
on the energies and momenta of different massive neutrinos, as
long as the relativistic dispersion relation $E_k^2=p_k^2+m_k^2$
is satisfied. This is why the standard derivation of the neutrino
oscillation probability gives the correct result, in spite of the
unrealistic equal momenta assumption.''

 We would like to point out that the replacing $t \simeq x = L$ is
fraught with at least two problems:
\begin{enumerate}
\item If one assumes that $x=t$, then the so-called space
velocities $\bar v = x/t$ of all three neutrinos are identical.
But in this case the plane wave approximation is not valid.
\item If, on the other hand, one considers the so-called
kinematical velocities $v_k = p_k/E_k$, according to the
plane-wave picture, then there appears a correction to the Eq.(5)
of ref. \cite{Giunti}:
\end{enumerate}
\begin{equation}
x = v_k t = \frac{p_k}{E_k}t \simeq (1-\frac{m_k^2}{2E_k^2})t.
\end{equation}
 Hence $t = (1+\frac{m_k^2}{2E_k^2})x$,
\begin{equation}
E_k t - p_k x \simeq (E_k - p_k)L +
\frac{m_k^2}{2E_k^2}L = 2\frac{m_k^2}{2E_k^2}L
\end{equation}
and one gets the standard formula multiplied by the notorious
``factor of 2'', which had been shown to be wrong. Thus, the
``theory'' of ref. \cite{Giunti} is false.

 In order to derive the standard phase in the plane wave approximation for
the ``clockless'' neutrino oscillation experiments\footnote{In all
neutrino oscillation experiments performed up to now time interval
is not measured, only the distance between the source and detector
is known. In that sense these experiments are ``clockless'' and
imply integration over $t$ which kills $t$-dependent terms.}, one
has to assume that all three $E_k$ are equal: $E_1=E_2=E_3=E$ (see
refs. \cite{Stodolsky}, \cite{Lipkin}) and replace each $p_k x$ by
$EL-\frac{m_k^2}{2E}L$.

\vspace{5mm}
\noindent
{\bf \large Acknowledgements}

\vspace{3mm}

This work was partly supported by the grants RFBR No.2328.2003.2,
No.01-02-17682a, grant INTAS No.99-0590 and by the A. von Humboldt
award to L.O.

\end{document}